\documentclass[aps,prb,graphicx,eqsecnum,
twocolumn,
floats,showpacs]{revtex4}
\usepackage[centertags]{amsmath}
\usepackage{amsfonts}
\usepackage{amssymb}
\usepackage{amsthm}
\usepackage[dvips]{graphicx}
\begin{document}
\title{Multistage Kondo effect as a manifestation of dynamical symmetries
in the single- and two-electron tunneling}

\author{K. Kikoin}
\affiliation{Raymond and Beverly Sackler Faculty of Exact
Sciences, School of Physics and Astronomy, Tel-Aviv University,
Tel-Aviv 69978 Israel}

\begin{abstract}
The concept of dynamical symmetries is used for formulation of the renormalization
group approach to the Kondo effect in the Anderson model with repulsive and attractive 
interaction $U$. It is shown that the generic local symmetry of the Anderson
Hamiltonian is determined by the $SU(4)$ Lie group. The Anderson Hamiltonian is 
rewritten in terms of the Gell-Mann matrices of the 4-th rank, which form the 
set of group generators and the basis for construction of irreducible 
vector operators describing the excitation spectra in the charge and spin sectors.
The multistage Kondo sceening is described in terms of the local $SU(4)$ dynamical 
symmetry.
 It is shown that the similarity 
between the conventional Kondo cotunneling
effect for spin 1/2 in the positive $U$ model and the Kondo resonance for pair tunneling 
in the negative $U$ model is a direct manifestation of implicit $SU(4)$ symmetry of
the Anderson/Kondo model.  
\end{abstract}

\maketitle

\section{Introductory remarks}\label{IIIintro}

In the 60-es and 70-es, when the basic concepts of dynamical
symmetries have been formulated and elaborated,
\cite{Nee62,GM64,Zweig64,NeGm,Barut64,DGN65} only few physical
realizations of these symmetries could be found in the realm of
existing quantum mechanical objects (see \cite{Engle,MalMan79} for a
review). On may mention the hydrogen atom
\cite{MORS65,SMOR65,MalMan65} and the harmonic oscillator in various
spatial dimensions \cite{GoLi59,Barut65b,Hwa66} as the systems for
which the study of their dynamical symmetries revealed additional
facets of excitation spectra and response to external fields. Rapid
progress in the nanotechnology and nanophysics during two recent decades
significantly extended the field of applicability of these concepts
and enriched the theory with some new ideas.

Contemporary nanophysics
\cite{CowAus01,SaCi03,CuFaR05,Nat06,Hanson07} deals with the
artificial structures which consist of finite number of electrons
confined within a tiny region of space, where the energy spectrum of
electrons is discrete. As a result, such objects can be treated as
"zero-dimensional" artificial atoms or molecules with spatially
quantized discrete states, well defined symmetry and controllable
electron occupation. Besides, modern technologies allow fabrication
of devices where a "natural" atom or molecule is spatially isolated
from the rest part of a device, so that the physical properties of
an {\it individual} atom or atomic cluster may be studied
experimentally.

In this paper we analyze the dynamical
 symmetries which arise when a group theoretical approach is used in the description of a
 contact between a few electron nanosystem ${\cal S}$ with definite
 symmetry $\sf{G}_{\cal S}$
 and a macroscopic system ${\cal B}$ ("bath" or "reservoir"). Due to this contact
 the symmetries of the system ${\cal S}$ and the corresponding
 conservation laws are violated. If the contact between two
 systems is weak enough, the dynamics of interaction may be
 described in terms of transitions between the eigenstates of a
 system ${\cal S}$ belonging to different irreducible
 representations of the group $\sf{G}_{\cal S}$ generated
 by the operators which obey the algebra $\sf{g}^{}_{\cal S}$.
 If the operators
 describing transitions between these eigenstates
 together with generators of the group $\sf{G}_{\cal S}$
 form the enveloping algebra $\sf{d}^{}_{\cal S}$ for the algebra
 $\sf{g}^{}_{\cal S}$, one may say
 that the system ${\cal S}$ possesses a dynamical symmetry
 characterized by some group $\sf{D}_{\cal S}$. The
 dynamical symmetry group technique offers mathematical tools for the unified approach to
 quantum objects, which allows one to consider not only the
 spectrum of a system ${\cal S}$, but also its response to external
 perturbation violating the symmetry $\sf{G}_{\cal S}$ and various
 complex
 many-body effects characterizing the interaction between the system ${\cal
 S}$ and its environment ${\cal B}$. In this paper we discuss a general 
algorithm of dynamical symmetry group approach to the few electron objects 
${\cal B + S}$ and its practical application to the single-electron
tunneling in complex quantum dots and single molecule transistors.
This tunneling is described in a framework of the Anderson model
with repulsive and attractive interaction between the confined
electrons.

\section{Hubbard operators generating the spectrum of nanoobject}

 Following the definition used in Ref. \onlinecite{MalMan65}, we
define the dynamical symmetry group ${\sf D}^{}_{\cal S}$ as a Lie
group of finite dimension characterized by the irreducible
representations which act in the whole Hilbert space of eigenstates
$|l\lambda\rangle$ of the Schroedinger equation
\begin{equation}\label{2.1}
\hat H |l\lambda\rangle = E_l|l\lambda\rangle
\end{equation}
describing the quantum system ${\cal S}$. Here $l$ is the index of
irreducible representation and $\lambda$ enumerates the lines of
this representation.  The projection operators
\begin{equation}\label{2.2}
X_{(l)}^{\lambda\mu} = |l\lambda\rangle \langle l\mu|
\end{equation}
play the central part in the procedure of construction of
irreducible representations $l$ of the group of Schroedinger
equation ${\sf G}^{}_{\cal S}$.  The basic property of these
operators is given by the equation
\begin{equation}\label{2.3}
X_{(l)}^{\lambda\mu}|l'\nu\rangle =
\delta_{ll'}\delta_{\mu\nu}|l\lambda\rangle\,.
\end{equation}
One may add to the set (\ref{2.2}) the operators
\begin{equation}\label{2.4}
X_{(ll')}^{\lambda\mu} = |l\lambda\rangle \langle l'\mu|
\end{equation}
which project the states belonging to different irreducible
representations $(l\neq l')$ of the group ${\sf G^{}_S}$ one onto
another.
 These operator may be also used for construction of the Lie algebras
 ${\sf d}^{}_{\cal S}$
generating the spectrum of eigenstates of the Schroedinger equation and
transitions between these states. 
 Unifying the notations $|l\lambda\rangle
=\Lambda\rangle$, we obtain the commutation relations
\begin{equation}\label{2.1a}
 [X^{\Lambda\Lambda'},\hat H] = (E_{\Lambda'} - E_\Lambda)\hat H
\end{equation}
The right hand side of Eq. (\ref{2.1a}) turns into zero provided the
states $\Lambda$ and $\Lambda'$ belong to the same irreducible
representation of the group ${\sf G}^{}_{\cal S}$.

  If a closed algebra ${\sf
d}^{}_{\cal S}$  exists for the set of operators (\ref{2.2}), (\ref{2.4}),
then one may state that the system described by the
Hamiltonian (\ref{2.1}) possesses the dynamical symmetry ${\sf
D}^{}_{\cal S}$. This algebra is conditioned by the norm
\begin{equation}\label{2.1c}
 \sum_\Lambda  X^{\Lambda\Lambda} = 1
\end{equation}
and by the commutation relations for the operators
$X^{\Lambda_1\Lambda_2}$.
  In the general case these relations
may be presented in the following form \cite{Hub2}
\begin{equation}\label{2.1b}
 [X^{\Lambda_1\Lambda_2},X^{\Lambda_3\Lambda_4}]^{}_{\mp}=
X^{\Lambda_1\Lambda_4}\delta_{\Lambda_2\Lambda_3}\mp
 X^{\Lambda_3\Lambda_3}\delta_{\Lambda_1\Lambda_4}
\end{equation}
The ``general case'' implies that the Fock space includes the states which
may belong to different charge sectors, i.e. changing the state
$\Lambda_1$ for the state $\Lambda_2$ means changing the number of
fermions  ${\cal N}_{\Lambda_2} \to {\cal N}_{\Lambda_1}$ in the many-particle system. 
If both  ${\cal N}_{\Lambda_1} - {\cal N}_{\Lambda_2}$ and 
${\cal N}_{\Lambda_3}-{\cal N}_{\Lambda_4}$ are odd
numbers (Fermi-type operators), the plus sign should be chosen in
Eq. (\ref{2.1b}). If at least one of these differences is zero or
even number (Bose-type operators), one should take the minus sign.

The operators $X^{\Lambda_1\Lambda_2}$ were exploited by J. Hubbard as a
convenient tool for description of elementary excitations in
strongly correlated electron systems (SCES). His seminal model of
interacting electron motion in a narrow band, known now as the
Hubbard model \cite{Hub1,Hub2,Hub3} was the first microscopic
model of SCES for which the conventional perturbative approach
based on the Landau Fermi liquid hypothesis turned out to fail. Now
the realm of SCES is really vast, and the most of artificial
nanostructures in fact belong to this realm. In particular, complex
quantum dots under strong Coulomb blockade are typical examples of
short Hubbard chains or rings.

The Hubbard operators (\ref{2.4}) obeying the commutation relation
(\ref{2.1a}) provide a convenient tool for construction of the algebras
generating the dynamical symmetry groups of the Schroedinger operator
$(\hat H-E)$ or theresolvent operator
$\hat {\sf R} = (\hat H-E)^{-1}$.  These operators may be used for 
construction of the generators of the appropriate group ${\sf
D}^{}_{\cal S}$ and
 irreducible tensor operators ${\cal O}^{(r)}$
(scalars, $r=0$, vectors, $r=1$, and tensors $r=2,3\ldots$) which
transform along the representations of the group ${\sf
D}^{}_{\cal S}$: 
\begin{equation}\label{2.5} 
 {\cal O}^{(r)}_{\varrho} = \sum_{\Lambda\Lambda'}
 \langle \Lambda |{\cal O}^{(r)}_\varrho|\Lambda'\rangle X^{\Lambda\Lambda'}.
\end{equation}
Here the index $\varrho$ stands for the components of irreducible tensor
operator of the rank $r$. On the one hand, it is clear that the
operators $X^{\Lambda\Lambda'}$ generate all the
eigenstates of the Hamiltonian $\hat H$ from any given initial
state $\Lambda'$. The components of the operator
 ${\cal O}^{(r)}$ form their own closed algebra, which characterizes
the dynamical symmetry
group provided the Hamiltonian $\hat H$ possesses such symmetry.
Having in mind the application of this technique to the geometrically confined
nanoobjects, we restrict ourself by the discrete eigenstates.

The Clebsch-Gordan expansion (\ref{2.5}) is the basic equation
which allows one to treat the dynamical symmetries of nanoobjects
in a systematic way. The principal difference between the
dynamical symmetries of SCES and those of integrable models is
that in the latter case the spectrum of the object and its
dynamical symmetries are known exactly, while in the former case
as a rule only some part of excitation spectra (usually its lower
part)
 may be found analytically and
classified by symmetry. This means that one may judge about the
dynamical symmetry of the spectrum only within the definite energy
interval $\cal E$. Respectively, the characteristic energy scale
may be different for different problems. 

Our main subject in this paper is the
Kondo effect in quantum dots. \cite{Glara88,NgLe88} The hierarchy
of the energy scales in this problem is well known.
\cite{WTs81,AFL81} The Kondo effect arises as a result of
orthogonality catastrophe in the Anderson model,\cite{Anders67}
where the conduction electrons in the Fermi sea of metallic
electrodes play part of the subsystem ${\cal B}$ and the strongly
correlated electrons in the quantum dot represent the subsystem
${\cal S}$. The largest energy scales in the Anderson model are
the width of conduction band $D$ in the subsystem ${\cal B}$ and
the energy of Coulomb blockade $Q$ in the subsystem ${\cal S}$.
The electrons confined in the nanoobject (quantum dot) are
characterized by the ionization energy $\epsilon_i$.  Next in the
hierarchy of energies are the tunneling amplitude $V$ and the
tunneling rate $\Gamma=\pi\rho_0 V^2$ characterizing the process
of electron tunneling through the potential barrier, which
separates two subsystems. Here $\rho_0$ is the density of electron
states at the Fermi level $\varepsilon^{}_F$ of the electron
liquid in the leads. The Kondo effect arises in the single-electron
tunneling regime under the restrictions of strong Coulomb blockade
$Q$. In this regime the charge transport between the source and
drain electrodes constituting the subsystem $\cal B$ is realized
as the electron cotunneling, where an electron from the source may
tunnel into the dot $\cal S$ only provided another electron leaves
the dot for the drain. Cotunneling, which arises in the fourth
order in $V$, is characterized by the energy $J$. Finally, the
energy scale of Kondo effect $E_K\sim
\sqrt{D\Gamma}\exp(-1/\rho^{}_0J)$ characterizes the crossover
from the weak coupling regime $J\ll 1$ to the strong coupling,
where $J$ is enhanced due to the multiple creation of low-energy
electron-hole pairs in the leads in the process of cotunneling.
The Kondo energy also scales the excitations above the ground
"Kondo-singlet" state.\cite{WTs81,AFL81} The hierarchy of all
these energies is
\begin{equation}\label{2.6}
D,U \gg \epsilon_i\gg V \gg \Gamma \gg J \gg E_K
\end{equation}

An effective way to describe the crossover from the weak coupling
to the strong coupling Kondo regime is the renormalization group (RG)
approach.\cite{And70,AbMig70,FowZaw71,Jeff77,Hald78} In this
method the renormalization of parameters $\epsilon_i, \Gamma, J$
in (\ref{2.6}) in the course of reduction of the energy scale
$\cal E$ from  high energies $\sim D,U$ to low energies still
exceeding $E_K$ is calculated. Our purpose is to describe this
procedure in terms of dynamical symmetries \emph{which change in
the course of reduction of the energy scale} $\cal E$. It was noticed
that the multistage Kondo screening predetermines the non-universal features
of the Kondo tunneling in the quantum dots with even occupation.
\cite{GuiTaglia00,EtoNazar00,KA01} In that case the relevant dynamical symmetry
groups are $SO(n)$ with $n =4 - 8$.\cite{KA01,KKA04}
In this paper we will show that this
language is useful already in the studies of the "ordinary" Kondo
effect for quantum dots with odd electron occupation $ \cal N$
characterized by spin 1/2. The relevant Lie groups are $SU(n)$ with $n=3,4$.

\section{Dynamical symmetries in quantum dots}

As was mentioned above, the
dynamical symmetries of confined electrons in the quantum dot $\cal S$
are revealed in its interaction with the "Fermi bath"
$\cal B$ of conduction electrons, The Anderson Hamiltonian describing
the coupling between two subsystems reads
\begin{equation}\label{3.bd}
\hat H = \hat H_d + \hat H_b + \hat H_{db}
\end{equation}
where three terms describe the nanoobject, the Fermi-bath and
their coupling, respectively. The term $\hat H_{db}$ in
general case includes the direct coupling (quantum tunneling of
electrons between two subsystems), the direct interaction of
Coulomb and exchange nature and the indirect (kinematic)
interaction induced by the tunneling. If the symmetry of
nanoobject is well defined, the Hamiltonian $\hat H_d$ may be
diagonalized by means of projection operators (\ref{2.2}), and the
generators of dynamical symmetry group (\ref{2.5})  arise in the
interaction term $\hat H_{db}$ in combination with the operators
describing the excitations in the Fermi bath. These symmetries cannot
be treated in the same way as the symmetries of the integrable systems discussed in
the monographs \onlinecite{Engle,MalMan65} and the references therein,
because the interaction not only activates the symmetry ${\sf
D}_{\cal S}$ of the nanoobject but also involves the charge,
orbital and spin degrees of freedom of the bath. This principal
difference was pointed out in Refs. \onlinecite{KA01}, where the quantum
tunneling through an artificial molecule (double quantum dot) with
even electron occupation ${\cal N}=2$ in presence of the many-particle
interaction of Kondo type was described by means of the generators
of the  $SO(4)$ group.

To take the dynamical symmetries explicitly in the calculations of
excitation spectra and in the studies of spin and charge transport in
nanoobject, one should adhere the following paradigm:
\cite{KAKr04}
   \begin{itemize}
    \item When diagonalizing  $\hat H_d$ use the projection
    operators in accordance with Eqs. (\ref{2.1}) -- (\ref{2.1a});

    \item Construct the operators $X^{\Lambda\Lambda'}$, which describe
    transitions between all the states in the "supermultiplet" of
    eigenstates of $\hat H_d$ belonging both to the same and to
    different irreducible representations of the symmetry group ${\sf G}_{\cal S}$
    of the Hamiltonian $\hat H_d$ and determine the relevant closed algebra generating
    the dynamical symmetry group ${\sf D}_{\cal S}$;

    \item Rewrite $\hat H_{db}$ in terms of the configuration change
    operators (\ref{2.4}) belonging to adjacent charge sectors
    ${\cal N}\to {\cal N}\pm 1$;

    \item When projecting the original Anderson Hamiltonian
    (\ref{3.bd}) on the subspace of low-energy states
    $\langle\bar\Lambda|\ldots|\bar\Lambda\rangle$ by means of the
    Schrieffer-Wolff (SW) transformation \cite{SW66} or its generalizations,
    express the Hubbard operators which arise in this transformation
    via the generators of corresponding dynamical symmetry group using 
    expansion (\ref{2.5}).

   \end{itemize}

To demonstrate this paradigm in action, let us consider the
textbook example of a cell which may contain zero, one or two
electrons with zero orbital moment. The Hamiltonian of this toy
model
\begin{equation}\label{2Hub.2}
\hat H_d = \epsilon_d \sum_{\sigma=\uparrow,\downarrow}
d^\dag_{i\sigma}d^{}_{i\sigma} +
  Un_{id\uparrow}n_{id\downarrow}
\end{equation}
is nothing but the single-site Hamiltonian describing the
elementary cell of the non-degenerate Hubbard model \cite{Hub1}
 with variable occupation number ${\cal N}=0,1,2$
(``Hubbard atom''). Using definition (\ref{2.4}) of the
Hubbard operator, we rewrite $\hat H_d$ in the universal form
\begin{equation}\label{3.hub}
\hat H_d= \sum_\Lambda E_\Lambda X^{\Lambda\Lambda}
\end{equation}
where $\Lambda=0,\sigma,2$ and the energy levels $E_\Lambda$ are
\begin{equation}\label{2Hub.4}
 E_0 =0,~~ E_{\uparrow}=E_{\downarrow} =E_1\equiv \epsilon_d, ~~ E_2 = 2\epsilon_d+U.
\end{equation}

\begin{figure}[h]
\includegraphics[width=8.5cm,angle=0]{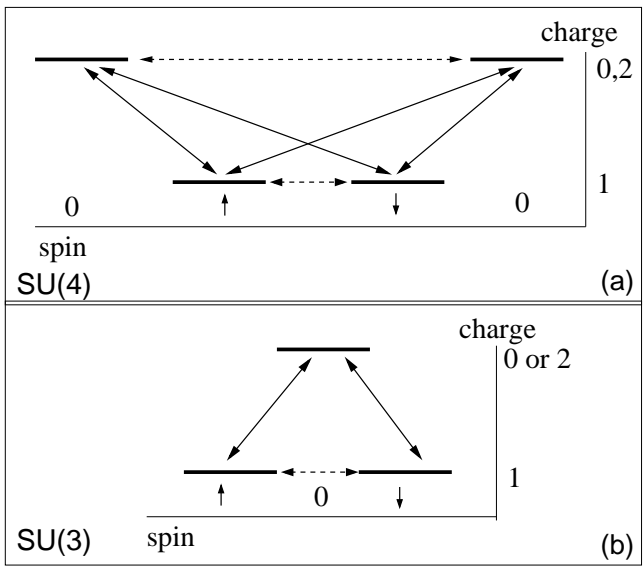}
\caption{
(a): Scheme of the energy levels for a Hubbard atom with the
$SU(4)$ dynamical symmetry describing transitions between the
states with occupation ${\cal N}= 0,1,2$.
(b) The same for a
reduced spectrum with the $SU(3)$ dynamical symmetry describing
transitions between the states with  ${\cal N}= 0~{\rm or}~2$ and
${\cal N}= 1$. The Bose-like transitions with even $\delta{\cal
N}=0,\pm2$ are marked by the dashed arrows, the Fermi-like transitions with
odd $\delta{\cal N}=\pm 1$ are marked by the solid arrows.
}
\label{Fig2_1}
\end{figure}

It is convenient to arrange the energy levels in accordance with the
available charge and spin sectors (Fig. \ref{Fig2_1}a). The arrows
connecting the levels $E_\Lambda$ and $E_{\Lambda'}$ correspond to the
Hubbard operator $X^{\Lambda\Lambda'}$ and its complex conjugate.
This scheme visualizes the Fermi-like and Bose-like operators (solid
and dashed lines, respectively), which obey the commutation
relations (\ref{2.1b}). There are 15 such operators forming a closed
superalgebra containing both commutators and anticommutators.
The Hubbard operators may be regrouped into the generators
of the $SU(4)$ group, known as the Gell-Mann matrices of 4th rank
$\lambda_1 - \lambda_{15}$ (see Appendix). Thus the generic
dynamical symmetry of Hubbard atom which is realized within the
energy interval ${\cal E} \sim U, D$ is $SU(4)$.

 Reduction of the energy scale to the interval $\epsilon_i < {\cal E} \ll
 U$ results in quenching the doubly occupied levels and
 corresponding reduction of the dynamical symmetry from $SU(4)$ to
 $SU(3)$ (Fig. \ref{Fig2_1}b). The algebra generating this group
 contains eight Gell-Mann matrices of the 3rd rank $\lambda_1 -
 \lambda_8$ and the same number of Hubbard operators. Relations between the
 matrix of Hubbard operators and the Gell-Mann matrices for this group
 are also presented in Appendix. Further
 reduction of the energy interval ${\cal E} \ll \epsilon_i$
 results in complete suppression of charged sectors ${\cal N} \neq
 1$, so that we are left only with spin states
 $\sigma=\uparrow,\downarrow$. In this limit the dynamical
 symmetry is the same as the symmetry of the Hubbard atom, and the
 corresponding Lie group is $SU(2)$.

 Mathematically, non-trivial dynamical symmetries are described by
 semisimple groups, possessing ideals. These ideals are formed by
 some triads of Gell-Mann matrices, e.g. matrices
 $\lambda_1, \lambda_2,\lambda_3$, which form a subgroup $SU(2)$
 of a group $SU(n)$. If the states in the Fock space for the Hubbard atom are
 ordered as
 \begin{equation}\label{X.su10}
\bar\Phi_4 =\left(
 \begin{array}{cccc}
 \uparrow ~  & \downarrow  ~ &  0 ~ & 2
 \end{array}
\right)
\end{equation}
then the first three Gell-Mann matrices $\lambda_1 - \lambda_3$ are
related to the spin states in the charge sector ${\cal N} =1$ (see
Appendix).

It is expedient to rewrite the original Hamiltonian (\ref{3.hub})
in terms of the generators of the group $SU(4)$ in the case where all
four eigenstates (\ref{2Hub.4}) shown in Fig.\ref{Fig2_1}a  are
taken into account, and in terms of the $SU(3)$ generators in the case
when the polar states with ${\cal N}=2$ are frozen out
(Fig.\ref{Fig2_1}b). In the full space $\bar \Phi_4$ we obtain by means of
(\ref{X.xsu4})
\begin{eqnarray}\label{2.10}
\hat H^{SU(4)}_d &=&\frac{E_0}{4}\left(1-\frac{4}{\sqrt{3}}X_8\right)
\nonumber\\
&+&\frac{E^{}_1}{2}\left(1+\frac{2}{\sqrt{3}}X^{}_8
 +\frac{2}{\sqrt{6}}X^{}_{15}\right) \nonumber \\
& + &\frac{h}{2}X^{}_3
+ \frac{E_2}{4}\left(1- \sqrt{6}X^{}_{15}\right)
\end{eqnarray}
 Here the notation $X_\rho$ is used
for the Gell-Mann matrices $\lambda^{}_\rho$ defined in the Fock
space (\ref{X.su10}). The Zeeman term $hS^{}_z$ acting in the
charge sector ${\cal N}=1 $ is also added.  In the reduced Fock
subspace
 \begin{equation}\label{X.su11}
\bar\Phi_3 =\left(
 \begin{array}{cccc}
 \uparrow ~  & \downarrow  ~ &  0
 \end{array}
\right)~~ {\rm or}~~\left(
\begin{array}{cccc}
 \uparrow ~  & \downarrow  ~ &  2
 \end{array}
\right)
\end{equation}
the Hamiltonian of the Hubbard atom rewritten with the use of Eqs.
(\ref{X.xsu43}) acquires quite compact form
\begin{equation}
\hat
H^{SU(3)}_d=\frac{E_0}{3}\left(1-\sqrt{3}X_8\right)
+\frac{E^{}_1}{3}\left(1+\sqrt{3}X^{}_8
 \right) +\frac{h}{2}X^{}_3.
\end{equation}

The Hubbard atom is a minimal model which can be used for
description of a quantum dot with variable occupation ${\cal N}$
coupled with the bath by means of the tunneling channel. The
equilibrium occupation of the dot may be changed by means of
injection of an electron or a hole from the metallic reservoir.\cite{CowAus01}
This occupation fluctuates dynamically due to the single electron
tunneling (SET) between the dot and the leads. The Coulomb
blockade parameter $Q$ plays the same part as the Coulomb repulsion
$U$ in the original Hubbard model. In the general case of, say, planar
quantum dot the energy spectrum of a quantum dot contains many
discrete states without definite angular symmetry. Only the
highest occupied (HO) and the lowest unoccupied (LU) states are
involved in single electron tunneling through such quantum dot.
The Hamiltonian of subsystem $\cal S$ in the Hamiltonian
(\ref{3.bd}) has the form
\begin{equation}\label{3.tun}
\hat H_d = \sum_j \varepsilon_j d^\dag_{j\sigma}d^{}_{j\sigma}+
\hat H_{\rm int} + Q\left(n_{\rm dot} - \frac{v_gC_g}{e}\right)^2
\end{equation}
Here the index $j$ enumerates the levels bottom-up.  $\hat H_{\rm
int}$ is the electron-electron interaction in the quantum dot,
Usually the self-consistent Hartree term is included in the
definition of discrete levels $\varepsilon_j$, and the relevant
contribution to $\hat H_{\rm int}$ is the exchange between the
electrons occupying different levels of a \emph{neutral} quantum
dot. $Q =e^2/2C$ is the capacitive energy of the dot, $n_{\rm
dot}= \sum_{j\sigma}^{({\mbox \tiny
ext})}d^\dag_{j\sigma}d^{}_{j\sigma}$ is the number of {\it extra}
electrons or holes which are injected in the dot due to tunneling described
by the Hamiltonian $ \hat H_{db}$.
\begin{equation}
 \hat H_{db} = \sum_{l=s,d}\sum_{jk\sigma}\left(W_{lj}d^\dag_{j\sigma}
  c^{}_{lk\sigma} + {\rm H.c.}\right).
\end{equation}
Corrections to the capacitive
energy take into account the capacitance of the gate $C_g$ and the
gate voltage $v_g$. If the hierarchy of the energy scales
\begin{equation}\label{3.block}
Q > (\delta \varepsilon, J) \gg W_{lj}
\end{equation}
takes place ($\delta\varepsilon$ is the interlevel spacing between
the HO and LU states, $J$ is the exchange coupling constant), then
one may assert that the charge transfer through the quantum dot
occurs in the SET regime.
\begin{figure}[h]
\includegraphics[width=8.5cm,angle=0]{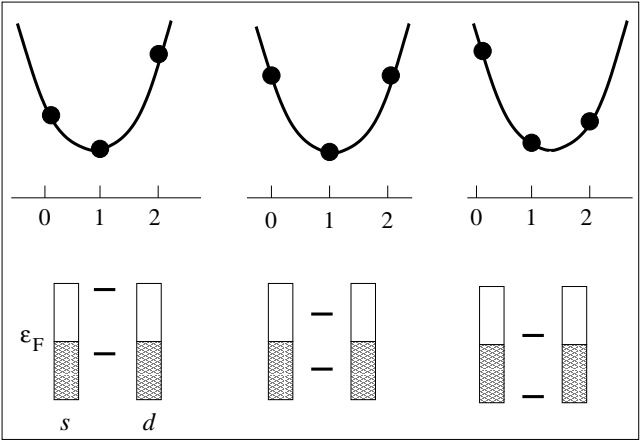}
\caption{Upper panel: Variation of the energy of the quantum dot
$E_{\rm el}({\cal N})$ as a function of the gate voltage. Lower panel:
corresponding variation of addition energies for electron and hole
excitations relative to the Fermi level in the leads. See the text for
further discussion. } \label{Fig2_2}
\end{figure}

Variation of the energy spectrum and the occupation of the quantum
dot as a function of a gate voltage is exemplified in Fig.
\ref{Fig2_2}. The "Hubbard parabolas" \cite{Hub1,Hub2,Hub3}
represent the energy $E_{\rm el}({\cal N})$ of the isolated
quantum dot with the Hamiltonian (\ref{3.tun}). Three subsequent
diagrams for the occupation ${\cal N}=1$ show the asymmetric
configurations with quenched zero and two electron occupation
(the side diagrams) and the configuration with particle-hole symmetry
(the middle diagram). The single particle excitations are the
addition and extraction energies $E({\cal N}) - E({\cal N}\mp 1)$
which should be compared with the chemical potential of the bath
(the Fermi energy) -- see the lower panel of Fig. \ref{Fig2_2}.

Following the above scheme, we express the Hubbard Hamiltonian  (\ref{2.10})
and the tunneling term $H_{db}$
\begin{equation}\label{htun}
H_{db}= \sum_{k\sigma}( t_k d^\dag_{k\sigma}c^{}_{k\sigma} + {\rm
H.c.})
\end{equation}
via the operators from the triads (\ref{X.triad15}) which are
connected with the original Hubbard operators $X^{\Lambda\Lambda'}$
acting in the space (\ref{X.su10}) by the following relations:
\begin{eqnarray}\label{2.11}
&&{\sf T}^+=X^{\uparrow\downarrow},~~
   {\sf T}^-=X^{\downarrow\uparrow},~~
   {\sf T}_z=X^{\uparrow\uparrow}-X^{\downarrow\downarrow}
   \nonumber\\
&& {\sf V}^+=X^{\uparrow 0},~~
   {\sf V}^-=X^{0\uparrow},~~
   {\sf V}_z=X^{\uparrow\uparrow}-X^{00}\nonumber\\
&& {\sf U}^+=X^{\downarrow 0},~~
   {\sf U}^-=X^{0\downarrow},~~
   {\sf U}_z=X^{\downarrow\downarrow}-X^{00}\nonumber\\
&& {\sf W}^+=X^{\uparrow 2},~~
   {\sf W}^-=X^{2\uparrow},~~
   {\sf W}_z=X^{\uparrow\uparrow}-X^{22}\nonumber\\
&& {\sf Y}^+=X^{\downarrow 2},~~
   {\sf Y}^-=X^{2\downarrow},~~
   {\sf Y}_z=X^{\downarrow\downarrow}-X^{22}\nonumber\\
&& {\sf Z}^+=X^{02},~~
   {\sf Z}^-=X^{20},~~
   {\sf Z}_z=X^{00}-X^{22}\
\end{eqnarray}
Equations (\ref{2.11}) realize the general expansion scheme
(\ref{2.5}) for the irreducible \textit{vector} operators in the group $SU(4)$.
The triad $\vec{\sf T}$ is nothing but the set of spin 1/2
operators $(S^+,S^-,2S_z, )$ acting in the charge sector ${\cal
N}=1$. The triad $\vec{\sf Z}$ describes the two-particle excitations
$({\cal N}=0 \leftrightarrow {\cal N}=2)$. The rest four triads
describe transitions between different charge sectors $({\cal N}=1
\leftrightarrow {\cal N}=0,2)$. These operators enter the Anderson
Hamiltonian corresponding to the Hubbard parabolas of Fig.
\ref{Fig2_2}.

The Hamiltonian $\hat H_d^{SU(n)}$ may be expressed via the $z$-components of
irreducible vectors (\ref{2.11}) by means of the following relations
\begin{eqnarray}\label{qupe}
 {\sf P}& = & {\sf V_z} + {\sf U_z} = X^{11} - 2X^{00},\nonumber \\
 {\sf Q}& = & {\sf W_z} + {\sf Y_z} = X^{11} - 2X^{22}, \nonumber\\
 {\sf R}& = & {\sf P}-{\sf Q} =2(X^{22}-X^{00})
\end{eqnarray}
and the completeness condition (\ref{2.1c}) which in this case reads 
\begin{equation}\label{comple}
 X^{00}+X^{11}+ X^{22}=1, ~~~X^{11}=\sum_\sigma X^{\sigma\sigma} 
\end{equation}
[see also
Eq. (\ref{apm})].
We find from Eqs. (\ref{qupe}) and (\ref{comple}):
\begin{eqnarray}\label{hubbop4}
 X^{00}= \frac{1}{4}-\frac{1}{8}(3{\sf P}- {\sf Q}) \nonumber\\
 X^{22}= \frac{1}{4}+\frac{1}{8}( {\sf P}- 3{\sf Q})\nonumber\\
 X^{11}= \frac{1}{2}+ \frac{1}{4}( {\sf P} + {\sf Q}).
\end{eqnarray}
Then the general $SU(4)$ configurations (the first and the 
third parabolas in Fig. \ref{Fig2_2})
are described by the Hamiltonian
\begin{eqnarray}
 \hat H^{SU(4)}&=&\frac{2E_1 +E_0+E_2}{4}\cdot{\sf 1} +
 \frac{h}{2}\cdot{\sf T}_z \nonumber\\
&+&\frac{E_{10}}{4}\cdot{\sf P} +
  \frac{E_{12}}{4}\cdot {\sf Q} + \frac{E_{20}}{8}\cdot {\sf R}
\end{eqnarray}
where $\sf 1$ is the unit matrix in the Fock space $\bar \Phi_4$, $E_{ij}=E_i-E_j$
are the addition/extraction energies.
Thus the
 operators ${\sf P}/4$, ${\sf Q}/4$, ${\sf R}/8$ and ${\sf T}_z/2$ describe
all Fermi- and Bose-like excitations shown in Fig. \ref{Fig2_1}a.
In the degenerate case $E_0=E_2\equiv E_e$ (second parabola in  Fig. \ref{Fig2_2})
this Hamiltonian reduces to
\begin{equation}
\hat
H^{SU(4)}_d=\frac{E^{}_{1}+E^{}_p}{2}\cdot{\sf 1} +\frac{h}{2}\cdot{\sf T}_z
+\frac{E^{}_{1e}}{4}\cdot
({\sf Q}+{\sf P})
\end{equation}

The tunneling term $\hat H_{db}$ also may be expressed via the generators
of $SU(4)$ group, namely via the ladder operators:
\begin{equation}
 \hat H_{db}^{SU(4)} = \sum_k t_k({\sf V}^\dag + {\sf W}^{\dag})c^{}_{k\uparrow}+
({\sf U}^\dag - {\sf Y}^-) c^{}_{k\downarrow} + {\rm H.c.}
\end{equation}

In the strongly asymmetric situations (the side configurations in Fig. \ref{Fig2_2},
where the excitation $E_{01}$ is
soft, whereas the excitation $E_{21}$ is frozen out or v.v.),
the symmetry of the dot is reduced from $SU(4)$
to $SU(3)$. Respectively, the system (\ref{hubbop4}) reduces to
\begin{eqnarray}\label{hubbop3}
 X^{00} = \frac{1}{3} - \frac{\sf P}{3},~~
 X^{11} = \frac{2}{3} + \frac{\sf P}{3},
\end{eqnarray}
 or, in terms of operators ${\sf U}, {\sf V}$
\begin{eqnarray}\label{hubbop3a}
 X^{\uparrow\uparrow} = \frac{1}{3} +\frac{2{\sf V}_z - {\sf U}_z}{3},~~
 X^{\downarrow\downarrow} = \frac{1}{3} +\frac{2{\sf U}_z - {\sf V}_z}{3}.
\end{eqnarray}

 The Anderson Hamiltonian acting in the space $\bar\Phi_3$ has the form
\begin{equation}\label{Hdsu3}
\hat H^{SU(3)}_d=\frac{2E_1+E_0}{3}\cdot{\rm 1}+\frac{E_{10}}{3}\cdot(
{\sf U}_z + {\sf V}_z) +\frac{h}{2}\cdot{\sf T}_z
\end{equation}
\begin{equation}\label{Htsu3}
\hat H^{SU(3)}_{db}=\sum_k t_k\left[\left({\sf
V}^+c_{k\uparrow}+{\sf U}^+c_{k\downarrow} \right) + {\rm
H.c.}\right].
\end{equation}
Thus the operators describing the charge Hubbard excitations in $SU(3)$
subspace $\bar \Phi_3$ (\ref{X.su11}) are
$\vec{\sf U}, \vec{\sf V}$, whereas the spin excitations are described by
the conventional spin
operator $\vec{\sf S} = \vec{\sf T}/2$.

The dynamics of charge and spin excitations in this case is
predetermined by the commutation relations for the group generators.
The operators ${\sf O}$ belonging to the same subgroup (triad)
commute in accordance with the standard $SU(2)$ relations
\begin{equation}\label{3.comA}
[{\sf O}_z,{\sf O}^{\pm}]= \pm 2{\sf O}^\pm,~~[{\sf O}^+,{\sf
O}^-]={\sf O}_z.
\end{equation}
The non-zero commutation relations between the operators belonging
to different triads ensure complex dynamical properties of
Hubbard-like SCES.
\begin{eqnarray}\label{2.13}
&&[{\sf U}^\pm,{\sf V^\mp}]=\pm{\sf T^\mp}, ~[{\sf U}^\pm,{\sf
V_z}]=\mp {\sf U^\pm}, \nonumber\\
&& [{\sf U}_z,{\sf V}^\pm]=\pm V^\pm,~[{\sf
U}_z,{\sf V}_z]=0.
\end{eqnarray}
Respectively, the non-zero anticommutation relations are
\begin{eqnarray}\label{2.13a}
&& \{{\sf U^+},{\sf U^-}\} = \frac{2 +{\sf V}_z- 2{\sf U_z}}{3}\nonumber \\
&& \{{\sf V^+},{\sf V^-}\} = \frac{2 +{\sf U}_z- 2{\sf V_z}}{3}
\end{eqnarray}

Then the excitations in the charge sector are described by the
Green functions, which may be found directly from equations of
motion for the generators of $SU(3)$ group.
\begin{equation}\label{2.14}
 G_v = \langle\langle{\sf V}^-(t){\sf V}^+(0)\rangle\rangle,~~
G_u = \langle\langle{\sf U}^-(t){\sf U}^+(0)\rangle\rangle,~~
\end{equation}
Respectively,
the excitations in the spin sector are given by the Green functions
\begin{equation}\label{2.15}
 G_s = \langle\langle{\sf S}^-(t)\cdot{\sf S}^+(0)\rangle,
\end{equation}
Here the double brackets stand for thermal averaging and
time-ordering operations specified for the retarded, advanced or
causal Green function. These functions can be easily found in the
atomic limit where only the term $\hat H_d^{SU(3)}$ is retained.
Solving equation of motion for the "Fermi-like" Green functions
which describe excitations in the charge sector, one gets by means
of the commutation and anticommutation relations (\ref{2.13}) and
(\ref{2.13a}):
\begin{eqnarray}\label{2.16}
&&G_v(\omega) = \frac{i}{2\pi}\frac{(2+\langle{\sf V}_z\rangle -
2\langle{\sf U}_z\rangle)/3}{\omega - \epsilon_d},\nonumber\\
&& G_u(\omega) =
\frac{i}{2\pi}\frac{(2+\langle{\sf U}_z\rangle - 2\langle{\sf
V}_z\rangle)/3}{\omega - \epsilon_d} .
\end{eqnarray}
Using the definitions (\ref{hubbop3}) and (\ref{hubbop3a}), we see
that the numerators in the Green functions (\ref{2.16}) are
nothing but the averages $\langle X^{00}\rangle+\langle
X^{\uparrow\uparrow}\rangle$ and $\langle X^{00}\rangle+\langle
X^{\downarrow\downarrow}\rangle$, so that these functions are
indeed the atomic Green functions for the Hubbard model
\cite{Hub3} rewritten in terms of the generators of the $SU(3)$ group.

The "Bose-like" Green function $G_s$ which describes the excitations
in the spin sector in theatomic limit has the usual form
\begin{equation}
G_s = \frac{i}{2\pi}\frac{\langle {\sf S}_z\rangle}{\omega - h}.
\end{equation}

\subsection{Three-fold way for the Hubbard atom}

As was noticed in the sixties, \cite{NeGm} various families of
hadrons are classified in accordance with the irreducible
representations of $SU(3)$ group (see also Ref. \onlinecite{ElDaub}). In
particular, 18 baryons form two multiplets corresponding to
representations $D^{(11)}$ (the octet of baryons with spin 1/2) and
$D^{(30)}$ (the decuplet of baryons with spin 3/2). The octet of spinless
mesons also transforms along the representation $D^{(11)}$. The
higher representations of the $SU(3)$ group are realized in the physics of
strong interaction because these "composite" particles possess
not only the spin and charge but also the isospin and hypercharge quantum
numbers, and the $SU(3)$ symmetry characterizes the latter variables. 
The elementary particles obeying the $SU(3)$
symmetry are the colored quarks. The $SU(3)$ symmetry in the hadron multiplets
under strong interaction is satisfied only approximately due to existence of 
electro-weak interaction, so that this symmetry may be treated as a
dynamical symmetry in the original sense of this notion.

The Hubbard atom with frozen doubly occupied states possesses only
two quantum numbers, namely spin and charge. Therefore, the
multiplet of Hubbard states is described by the lowest irreducible
representation $D^{(10)}$ of the $SU(3)$ group. To construct this representation, 
one
should recollect that the two of eight Gell-Mann matrices can be
diagonalized simultaneously. Following Ref. \onlinecite{ElDaub}, we choose
the representation with diagonal matrices ${\sf T}_z$ and ${\sf
Q}$. Then the set of allowed states is defined by two integer
numbers $\lambda,\mu$ so that the eigenstates are determined as
\begin{equation}\label{2.12}
M^{}_T = \lambda+\mu,~~ M^{}_Q = \frac{1}{3}(\lambda-\mu).
\end{equation}
The whole set of eigenstates form a two-dimensional triangular
lattice on the plane $(M^{}_T, M^{}_Q)$. Each irreducible
representation $D^{\bar\lambda\bar\mu}$ is marked by the indices
$\bar\lambda,\bar\mu$ corresponding to the state with the maximum
eigenvalue $\bar M^{}_Q$ and the maximum value of $\bar M^{}_T$
possible at this $\bar M^{}_Q$. Then the rest states forming this
irreducible representation are constructed by means of the ladder
operators ${\sf T}^\pm, {\sf U}^\pm, {\sf V}^\pm$ acting on the
state $|\bar M^{}_Q,\bar M^{}_T\rangle$.

\begin{figure}[h]
\includegraphics[width=6cm,angle=0]{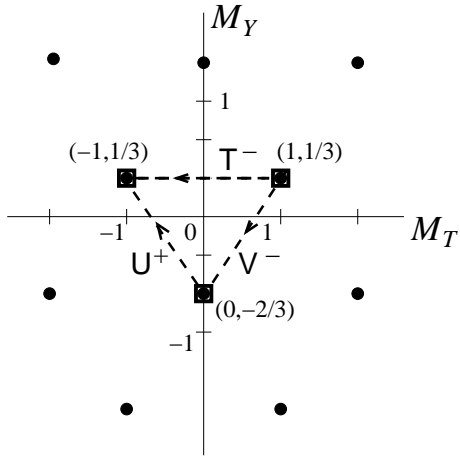}
\caption{Irreducible representation $D^{(10)}$ for the set
$\bar\Phi_3$} \label{Fig2_3}
\end{figure}

This procedure results in construction of the stars of basis
vectors $\vec D^{\lambda\mu}$ and the polygons connecting the points
generated by the ladder operators subsequently acting on the point
$(\bar M^{}_Q,\bar M^{}_T)$. In the case of baryon family the
corresponding multiplets are the hexagon with doubly degenerate cental
point for representation $D^{(11)}$ and the triangle with ten point in
its vertices and on its sides for representation $D^{(30)}$. In the
case of Hubbard atom the multiplet is represented by a triangle (Fig.
\ref{Fig2_3}) labeled in accordance with the state with the highest
quantum numbers $\lambda=1,\mu=0$, which corresponds to the state
$|{\cal N},\sigma\rangle = |1,\uparrow\rangle$ of the Hubbard
atom.
Two remaining components of the multiplet $\bar\Phi_3$ may be generated
from the state $|1,\uparrow\rangle$ by means of the ladder
operators ${\sf T}^- = X^{\downarrow\uparrow}$ and ${\sf
V^-}=X^{0\uparrow}$. First of these operators corresponds to the
"Bose-like" excitation with spin 1, and the second one is the
"Fermi-like" excitation with spin 1/2. The triangle $D^{(10)}$ is
closed by means of the operator ${\sf U^+}=X^{\downarrow 0}$. The
interrelations between the values of the parameters $\lambda,\mu$,
the eigenvalues of the operators ${\sf T}_z$ and $\sf Q$, and the
eigenvalues $|\Lambda\rangle$ of the Hubbard Hamiltonian are
presented in the following table
\begin{eqnarray}\label{2.table}
\begin{tabular}{|c|c|c|c|c|}
  \hline
   $\lambda$ & $\mu$ & $M_T$ & $M_Q$ & $\Lambda$ \\
  \hline
  1 & 0 & 1 & 1/3 & u \\
  0 & -1 & -1 & 1/3 & d \\
  -1 & 1 & 0 & -2/3 & h \\
  \hline
\end{tabular}
\end{eqnarray}
Here the notations ${\rm u,d,h}$ are used for the spin up, spin
down and hole states, respectively.

Thus we see that the dual nature of the Hubbard operators manifested
in the superalgebra with the commutation relations (\ref{2.1b})
allows one to use them
for construction of the generic $su(3)$ algebra formed by the spin and pseudospin
operators with the commutation relations (\ref{3.comA}), (\ref{2.13}).

Like in the case of baryons and mesons, this symmetry is violated
due to interaction with other subsystems. In our case this is the Fermi bath
$\cal B$. The source of this interaction is the tunneling coupling given
by the Hamiltonian $\hat H^{SU(3)}_{db}$. In analogy with the term
offered by Gell-Mann and Ne'eman for the multiplet of light
hadrons one may use the term "three-fold" (ternary) way for the SCES with
approximate $SU(3)$ symmetry. Generalization of this description for
the $SU(4)$ group is straightforward. In this case the phase space
for the irreducible representations is defined by the eigenvalues of the
operators $\sf P$, $\sf Q$, ${\sf T}_z$, and the lowest
irreducible representation of this group $D^{(100)}$ is represented by
a triangular pyramid in this 3D space.

  In the next section we will see how the
 hierarchy of dynamical symmetries of the Hubbard atom manifests
 itself in the RG evolution of the Anderson-Kondo problem.

\section{Two-stage renormalization group for $SU(3)$ and $SU(4)$ Anderson Hamiltonian}

The RG method is based on the idea of renormalization of model
parameters, which are relevant at low energy as a result
of the change of the scale of high energy excitations.\cite{FowZaw71} If the
model is renormalizable, any such parameter $P(\varepsilon)$ may
be represented as
\begin{equation}\label{scal}
P(\varepsilon)-P[(1+\kappa)\varepsilon]= - \kappa\varepsilon
P'(\varepsilon)
\end{equation}
where $\kappa$ is positive infinitesimal and the prime stands for the derivative. 
The quantity  $- \kappa\varepsilon
P'(\varepsilon)$ is the contribution to $P(\varepsilon)$ from the
high-energy states which are to be integrated out, preserving the
form of $P(\varepsilon)$ but changing its scale. Adopting
this approach, we immediately notice the inevitability of the three- or
two-stage RG procedure as a direct consequence of several energy scales
inherent in the Anderson model and the dynamical $SU(n)$ symmetry
of its excitation spectrum with \textit{n} = 4 or 3.

Taking as an example the first of three Hubbard parabolas in Fig. \ref{Fig2_2},
we see that in this case the
highest energy scale is the addition energy
${\cal E} \sim \epsilon_d + Q - \varepsilon_F$. The corresponding
 generators of $SU(4)$ group are $\vec{\sf W},\vec{\sf Y}$.
The next energy scale ${\cal E} \sim \varepsilon^{}_F -
\epsilon^{}_d$ is the extraction energy, and the relevant
generators are $\vec{\sf U}, \vec{\sf V}$. The lowest energy scale
${\cal E} \ll t^2/\epsilon_d$ is introduced by the second-order
cotunneling processes from the dot to the leads, which are
accompanied by the spin flips in the dot and creation of the
low-energy electron-hole pairs in the leads. The vector operator
$\vec{\sf T}$ is responsible for these processes. In other words,
we arrive to the the renowned Jefferson-Haldane-Anderson
renormalization group (RG) procedure. \cite{Jeff77,Hald78}  Basing
on the symmetry analysis of preceding section, the RG procedure
may be described in terms of the generators of $SU(4)$ group and
its subgroups with reduction of the symmetry $SU(4) \to SU(3) \to
SU(2)$ following the reduction of the energy scale $\cal E$.

Let us rederive for example  the scaling equations for the two stage 
RG $SU(3) \to SU(2)$
realized in the limit $U\to \infty$ in these new terms.
In the Anderson model the change of the energy scale in Eq.
(\ref{scal}) means the contraction of the electron bandwidth $D\to
D-\delta D$ in $\hat H_b$, The renormalized quantities are the
self energies $\Sigma_\eta(\varepsilon)$ of the Green functions
(\ref{2.14}) and (\ref{2.15}), $\eta=u,v,s$. The tunneling Hamiltonian 
(\ref{Htsu3}) gives the second-order
self-energy part for the Green
functions $G_v, G_u$ (\ref{2.16})
\begin{equation}
\epsilon_d = E_{10} +
\frac{\Gamma}{\pi}\int_0^D\frac{d\varepsilon}{E_{10}- \varepsilon}
\end{equation}
where $$\Gamma=\Gamma_u = \Gamma_v =\pi\sum_k
|t_k|^2\delta(\varepsilon_F -\varepsilon_k)
$$
is the spin-independent tunneling rate. The transformation
(\ref{scal}) results in the Jefferson-Haldane scaling equation
\cite{Jeff77,Hald78}
\begin{equation}
\frac{d\epsilon_d}{dD}= \frac{\Gamma}{\pi D}
\end{equation}
with the scaling invariant
\begin{equation}
\epsilon_d^* = \epsilon_d +\frac{\Gamma}{\pi}\ln\left(\frac{\pi
D}{\Gamma}\right).
\end{equation}
Thus, the evolution of the resonance level is determined by the
vectors $\vec{\sf U},\vec{\sf V}$ operating in the charge subsectors of the
group $SU(3)$. The same second-order processes generate the
 four-tail vertices $\sim {\sf V}^+{\sf
U}^-c^\dag_{k\downarrow}c^{}_{k'\uparrow}$, ${\sf U}^+{\sf
V}^-c^\dag_{k\uparrow}c^{}_{k'\downarrow}$ etc. Using the commutation
relations (\ref{2.13}), these vertices are combined in the conventional
Schrieffer-Wolff exchange interaction
\begin{equation}\label{2SW}
H_{\rm SW}^{}= J\vec{\sf S}\cdot \vec{\sf s}
\end{equation}
where $\vec{\sf s} = N^{-1}\sum_{kk'}c^\dag_{k'\sigma}\hat\tau c^{}_{k'\sigma'}$
is the local spin operator for conduction electrons, $\hat\tau$ is the the
vector of Pauli matrices
$J |\sim t^{}_{k_F}|^2/E_{10}$
is the indirect Kondo exchange. The scaling equations for this Hamiltonian
may be derived by means of the Anderson's RG procedure.

In the symmetric configuration (middle parabola in Fig. \ref{Fig2_2}) the
Jefferson -- Haldane -- Anderson scaling theory
is in fact the manifestation of reduction of the dynamical symmetry
$SU(4) \to SU(2)$,
in which the charge excitations represented by the vectors
$(\vec{\sf W}, \vec{\sf Y}, \vec{\sf U}, \vec{\sf V})$
are frozen out in the process of renormalization and the subgroup $\vec{\sf T}$ 
describing the spin degrees of freedom in the charge sector ${\cal N} =1$ 
represents the low-energy part of the spectrum responsible for the Kondo singularities. 
Five vectors of six triads available in the Gell-Mann set are involved in 
this two-stage procedure.

\begin{figure}[h]
\begin{center}
 \includegraphics[width=8.5cm,angle=0]{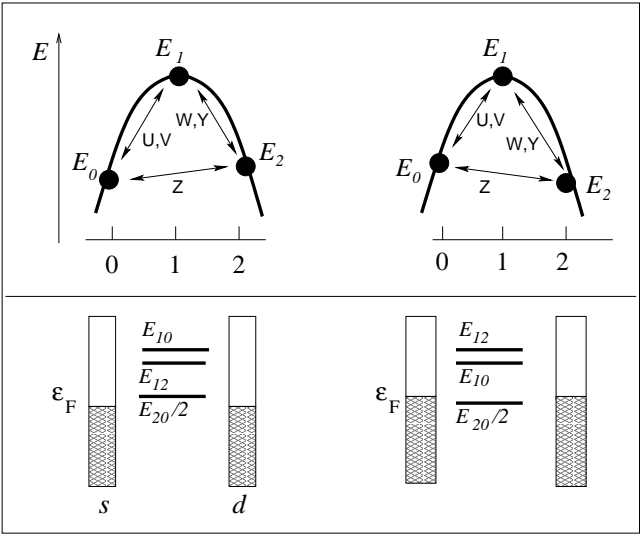}
 \caption{Upper panel: inverted Hubbard parabolas for the
negative $U$ Hubbard atom in the
cases of empty and doubly occupied shells. The interlevel transitions are
described by the operators generating the $SU(4)$ dynamical group.
  Lower panel: single-electron levels corresponding to the transitions
shown by the arrows in the upper panel (see the text for further explanation).}
\label{FigV_14}
\end{center}
\end{figure}

Another choice of 15 linearly independent generators of the
$SU(4)$ dynamical symmetry is possible in the case where instead
of the spin degeneracy of the ground state with ${\cal N}=1$ the
charge degeneracy of the two singlets with ${\cal N}=0,2$ is
realized. Such a possibility arises in the negative $U$ Anderson
model. \cite{And75,SchuF88,Taraph91} In this configuration the
vector $\vec{\sf T}$ is excluded from the renormalization procedure
due to quenching of the sector ${\cal N}=1$ at low energies. Instead
the vector $\vec{\sf Z}$ is involved in formation of the Kondo
singularities. The attractive interaction between the electrons in
the nanoobject in this model stems from the strong electron-phonon
interaction. Starting with the Anderson -- Holstein model where
the phonon subsystem is represented by the single Einstein mode
with the energy $\Omega_0$, one may perform the canonical
transformation,\cite{LaFi63} which transforms the electron-phonon
interaction into the polaron dressing exponent for the electron
tunneling rate, the polaron shift of discrete electron levels
and the phonon mediated electron-electron interaction. The latter
renormalizes the Hubbard interaction term in the Anderson
Hamiltonian
\begin{equation}\label{5.negu}
U' = U - 2 \lambda^2\Omega_0~ .
\end{equation}
Here $\lambda$ is the electron-phonon coupling constant. In the limit of strong
electron-phonon coupling the energy gain due to the phonon
mediated interaction overcomes the energy loss due to the Hubbard repulsion,
and one comes to the case $U' <0$. The negative $U$ model
may be realized in the single electron molecular transistors.
\cite{Cornag95a,Cornag95b,AlBrat03,Koch06,Koch07,LeWe09}
The interaction (\ref{5.negu}) should be
included in the term $\hat H_d$, so that in the negative $U$ case
 the Hubbard parabolas for the energy
spectrum are reversed relative to the usual shape shown in the middle 
configuration of Fig. \ref{Fig2_2}. 
The ``turned over'' diagrams corresponding to the two
nearly symmetric configurations shown in Fig. \ref{FigV_14}.
Like in the positive $U$ case, the transitions between the levels in the
Hubbard supermultiplet
are described by the operators  (\ref{2.11}) generating the $SU(4)$ dynamical
 symmetry group. We consider here
the configurations, where the singlet states $|\Lambda\rangle =|0\rangle,|2\rangle$
are degenerate or nearly degenerate, and the spin
doublet $|\Lambda\rangle = |\uparrow\rangle, |\downarrow\rangle$ is an excited virtual state
in the cotunneling processes. The two configurations presented in Fig. \ref{FigV_14}
 correspond to the
empty and completely filled two-electron shell of the Hubbard atom.
They are connected by the
particle-hole symmetry transformation, so it is enough to discuss one of them.

We will show below that the negative $U$ Anderson model may be
formally mapped on the positive $U$ model, by means of the
multistage RG method, which generalizes the Jefferson -- Haldane
-- Anderson procedure \cite{Jeff77,Hald78,And70} mentioned above.
In the positive $U$ case after freezing out the high-energy excitations
$E^{}_{01}$ and $E{}_{21}$  corresponding to injection
of a hole or of an electron into the singly occupied quantum dot at the
Jefferson-Haldane stage of the renormalization, one arrives at the Anderson stage 
of Kondo screening of spin excitations
in the sector ${\cal N}=1$ described by the vector operator $\vec{\sf T}$.
In the negative $U$ model
the spin excitations are exponentially suppressed from the very beginning.
After freezing out
the charge excitations $E^{}_{10}$ and $E^{}_{12}$ generated by the operators
$\vec{\sf U}, \vec{\sf V}, \vec{\sf W},\vec{\sf Y}$ 
we are left only with the two-particle charge excitations $E^{}_{20}$
generated by the operator $\vec{\sf Z}$.

Since the Jefferson-Haldane stage of the RG procedure is
realized exactly in the same way
as in the powitive $U$ Anderson model, we concentrate on the second stage, where the
$SU(4)$ dynamical symmetry group is reduced to
its $SU(2)$ subgroup represented by the triad $\vec{\sf Z}$. These operators act in
the subspace $\bar \Phi_2 =(0,2)$. The effective SW Hamiltonian in this subspace reads
\begin{equation}\label{5.cotun2}
\hat H_{\rm cotun} =N\frac{J_\perp}{2}\left( {\sf Z}^+{\sf B}^- 
+ {\sf Z}^-{\sf B}^+ \right)
 +NJ_\parallel  {\sf Z}_z {\sf B}_z~,
\end{equation}
where the components of the vector $\vec{\sf Z}$ are presented in the last 
line of the system (\ref{2.11})]. The components of the vector $\vec{\sf B}$ 
defined in the space of two-particle itinerant excitations are  
\begin{eqnarray}\label{5.bic}
 {\sf B}^+ &=& N^{-1}\sum_{kk'}c^\dag_{k\uparrow}c^\dag_{k'\downarrow},~~
 {\sf B}^- = N^{-1}\sum_{kk'}c^{}_{k\downarrow}c^{}_{k'\uparrow},\nonumber\\
 {\sf B}^{}_z &=& N^{-1}\sum_{kk'}\left(c^\dag_{k\uparrow}c^{}_{k'\uparrow} -
            c^{}_{k'\downarrow}c^\dag_{k\downarrow} \right) \nonumber\\
& = &
            N^{-1}\sum_{kk'}\sum_\sigma c^{\dag}_{k\sigma}c^{}_{k'\sigma} - 1
\end{eqnarray}
These operators obey the $su(2)$ commutation relations
\begin{equation}
 [{\sf B}^+,{\sf B}^-]= {\sf B}_z,~~ [{\sf B}^{}_z, {\sf B}^\pm]= \pm {\sf B}^\pm
\end{equation}
The transversal part of the Hamiltonian (\ref{5.cotun2}) describes the tunneling
of singlet electron pairs between the leads and the molecule, whereas its
longitudinal part stems from the band electron scattering on the charge fluctuations.

\begin{figure}[h]
\begin{center}
 \includegraphics[width=8.5cm,angle=0]{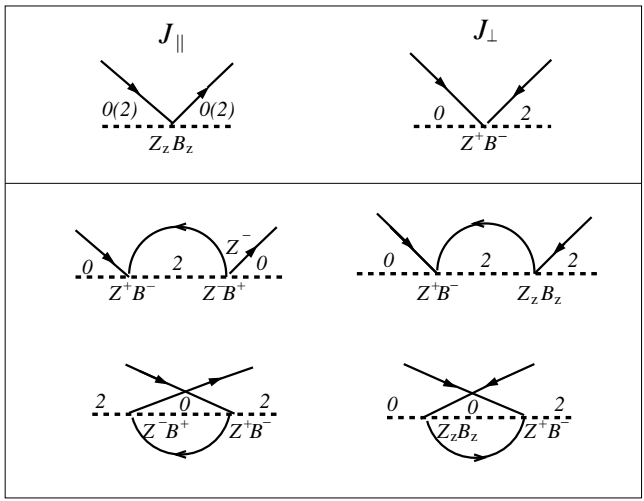}
 \caption{RG diagrams in the space $\bar\Phi_2 =(0,2)$. Upper panel: 
the bare vertices $J_{\perp}$ for the two-electron tunneling
and $J_{\parallel}$ for the charge scattering.
  Lower panel: the diagrams for the second-order renormalization of these vertices.
Solid lines stand for the conduction electron states, dashed lines
denote the charge states of the molecule.}
\label{FigV_13}
\end{center}
\end{figure}

Thus the Hamiltonian of two-electron tunneling is formally mapped onto the
anisotropic Kondo Hamiltonian \cite{SchuF88,Taraph91,Cornag95a} 
(see Fig.\ref{FigV_13}).
The origin of this anisotropy is the polaron dressing
of tunneling matrix elements. \cite{Cornag95a} This dressing is different for the
two-electron cotunneling and the electron scattering coupling parameters in
the Hamiltonian
(\ref{5.cotun2}). In the strong electron-phonon coupling limit,
$(\lambda/\Omega_0)^2 = S \gg 1$
\begin{equation}\label{5.anizex}
 \frac{J_\perp}{J_\parallel} =
\langle 2|0\rangle \sim e^{-2(\lambda/\Omega_0)^2}.
\end{equation}
The eventual source of this anisotropy is the overlap between the phonon
wave functions for a molecule in the charge states ${\cal N}=0$ and ${\cal N}=2$,
i.e. the Huang-Rhys factor $S$.
In a framework of the Anderson RG scalng procedure this
means that the renormalization diagrams for the two models are the same, 
namely the diagrams in the first and the second columns of Fig. \ref{FigV_13}
are mapped on the longitudinal and transversal components of the Kondo exchange
Hamiltonian.(\ref{2SW})
The mapping procedure implies the substitution
$\vec{\sf S} \to \vec{\sf Z}$, $\vec{\sf s} \to \vec{\sf B}$. 
The scaling equations, which follow
from these equations are the same as for the conventional anisotropic
Kondo model,\cite{And70} namely
\begin{eqnarray}
\frac{dj_\parallel}{d \eta} = - j_\perp^2, ~~
 \frac{dj_\perp}{d \eta} = - j_\perp j_\parallel
\end{eqnarray}
($j_i=\rho_0 J_i$).
In the case of strong anisotropy (\ref{5.anizex}) solution of this system
gives for the Kondo temperature the following equation \cite{Cornag95a}
\begin{equation}\label{5.tkondd}
 T_K \sim\left(\frac{j_\perp}{j_\parallel}\right)^{1/j_\parallel}
\sim \bar D \exp \left[-\frac{\pi\Omega_0}{2\Gamma}
 \left(\frac{\lambda}{\Omega_0}\right)^4 \right].
\end{equation}
The last equation in (\ref{5.tkondd}) is valid in the limit of
strong electron-phonon coupling $S\gg 1$.
Generally, the polaron narrowing of the tunneling rate results in a noticeable
decrease of $T_K$ in comparison with its value for the conventional Kondo effect.

\section{Concluding remarks}

In this paper we have shown that the ``atomic'' part $\hat H_d$ of the Anderson 
Hamiltonian may be treated as a four-level system, so that the interlevel transitions
induced by the tunneling term $\hat H_{db}$ activate the implicit $SU(4)$ dynamical 
symmetry of 
the model. The symmetry is the same for the negative and positive $U$ models. 
However, in spite of the formal similarity between the effective Hamiltonians for the
single electron cotunneling and the electron pair cotunneling, the background physics
is different in two versions of the Anderson model. In the positive $U$ Anderson model the tunneling in the middle
of the Coulomb window arises exclusively due to the many-body Abrikosov -- Suhl resonance.
In the negative $U$ model the resonance conditions for the two-electron tunneling arise
at $E_{02}=0$ irrelative to the many body particle-hole screening mechanism,
so that the zero bias anomaly in the tunneling conductance exists already at 
$T \gg T_K$, as well as the finite bias anomaly at $E_{02} \neq 0$.
\cite{Koch06,Koch07,LeWe09} One may say that the Anderson orthogonality catastrophe
\cite{Anders67} responsible for the many-body Kondo-like screening
at low $T$ only enhances the two-electron tunneling resonance already sharpened
due to non-orthogonality of the phonon clouds measured by the Huang -- Rhys factor
(\ref{5.anizex}). The finite difference $E_{02}\neq 0$
in the negative $U$ model
is equivalent to the finite magnetic field in the positive $U$ model: it results in the
appearance of two split finite bias peaks in the tunneling conductance.

Having in mind all these differences, one may state that the
multistage RG procedure reveals the hierarchy of reduced dynamical
symmetries $SU(4) \to SU(3) \to SU(2)$ in the Anderson model both
with the Hubbard repulsion for odd occupation and with the Hubbard
attraction for even occupation. In this paper we confined
ourselves with rephrazing the results already known in the Kondo
physics, but we believe that the use of operators generating the
set of the eigenstates of the zero Hamiltonian in the Anderson and
Hubbard models and thus revealing its internal symmetry may
facilitate the description of various dynamical properties of
strongly correlated electron systems. In particular, new
bosonization and fermionization procedures for the generators
(\ref{2.11}) alternative to the standard representations for the
$SU(n)$ groups \cite{ReNe83,Coleman84,AfMars88,KFO01} may be
elaborated. We will turn to these procedures in the forthcoming publications.

\appendix
\section{GELL-MANN MATRICES AND HUBBARD OPERATORS}

Here we summarize for the sake of convenience some properties of
the Gell-Mann matrices of 4th rank and their realization in the Hubbard
and Anderson models. The canonical form of these
matrices describing the symmetry of four-level systems is
\begin{widetext}
\begin{eqnarray}\label{X.su8}
&&\lambda_1 =
 \left(
 \begin{array}{cccc}
 0 & 1 & 0 & 0 \\
 1 & 0 & 0 & 0 \\
 0 & 0 & 0 & 0 \\
 0 & 0 & 0 & 0 \\
 \end{array}
\right),~~
 \lambda_2 =
 \left(
 \begin{array}{cccc}
 0 & -i & 0 & 0 \\
 i & 0 & 0  & 0\\
 0 & 0 & 0  & 0\\
 0 & 0 & 0 & 0 \\
 \end{array}
\right),~~
 \lambda_3 =
 \left(
 \begin{array}{cccc}
 1 & 0 & 0 & 0 \\
 0 & -1 & 0 & 0 \\
 0 & 0 & 0 & 0\\
 0 & 0 & 0 & 0 \\
 \end{array}
\right), \nonumber\\
&& \lambda_4 =
 \left(
 \begin{array}{cccc}
 0 & 0 & 1 & 0 \\
 0 & 0 & 0 & 0 \\
 1 & 0 & 0 & 0 \\
 0 & 0 & 0 & 0 \\
 \end{array}
\right)~~ \lambda_5 =
 \left(
 \begin{array}{cccc}
 0 & 0 & -i & 0 \\
 0 & 0 & 0 & 0 \\
 i & 0 & 0 & 0 \\
 0 & 0 & 0 & 0 \\
 \end{array}
\right),~~ \lambda_6 =
 \left(
 \begin{array}{cccc}
 0 & 0 & 0 & 0\\
 0 & 0 & 1 & 0\\
 0 & 1 & 0 & 0\\
 0 & 0 & 0 & 0\\
 \end{array}
\right), \\
&& \lambda_7 =
 \left(
 \begin{array}{cccc}
 0 & 0 & 0 & 0\\
 0 & 0 & -i & 0  \\
 0 & i & 0 & 0 \\
 0 & 0 & 0 & 0 \\
 \end{array}
\right),~~ \lambda_8 = \frac{1}{\sqrt{3}}
 \left(
 \begin{array}{cccc}
 1 & 0 & 0 & 0\\
 0 & 1 & 0 & 0\\
 0 & 0 & -2 & 0\\
 0 & 0 & 0 & 0 \\
 \end{array}
\right), ~~ \lambda_9 =
 \left(
 \begin{array}{cccc}
 0 & 0 & 0 & 1\\
 0 & 0 & 0 & 0  \\
 0 & 0 & 0 & 0 \\
 1 & 0 & 0 & 0 \\
 \end{array}
\right),\nonumber \\
&& \lambda_{10} =
 \left(
 \begin{array}{cccc}
 0 & 0 & 0 & -i\\
 0 & 0 & 0 & 0  \\
 0 & 0 & 0 & 0 \\
 i & 0 & 0 & 0 \\
 \end{array}
\right), ~~ \lambda_{11} =
 \left(
 \begin{array}{cccc}
 0 & 0 & 0 & 0\\
 0 & 0 & 0 & 1  \\
 0 & 0 & 0 & 0 \\
 0 & 1 & 0 & 0 \\
 \end{array}
\right),~~ \lambda_{12}=
 \left(
 \begin{array}{cccc}
 0 & 0 & 0 & 0\\
 0 & 0 & 0 & -i  \\
 0 & 0 & 0 & 0 \\
 0 & i & 0 & 0 \\
 \end{array}
\right), \nonumber  \\
&& \lambda_{13} =
 \left(
 \begin{array}{cccc}
 0 & 0 & 0 & 0\\
 0 & 0 & 0 & 0  \\
 0 & 0 & 0 & 1 \\
 0 & 0 & 1 & 0 \\
 \end{array}
\right), ~~ \lambda_{14} =
 \left(
 \begin{array}{cccc}
 0 & 0 & 0 & 0\\
 0 & 0 & 0 & 0  \\
 0 & 0 & 0 & -i \\
 0 & 0 & i & 0 \\
 \end{array}
\right), ~~ \lambda_{15} = \frac{1}{\sqrt{6}}
 \left(
 \begin{array}{cccc}
 1 & 0 & 0 & 0 \\
 0 & 1 & 0 & 0 \\
 0 & 0 & 1 & 0 \\
 0 & 0 & 0 & -3 \\
 \end{array}
\right), \nonumber
\end{eqnarray}
\end{widetext}
(see, e.g. \cite{tilmaa}).
First eight matrices contain the 3-rd rank Gell-Mann operators of the
$SU(3)$ group as submatrices. The operators $\lambda_9 -
\lambda_{15}$ generate transitions between the triplet and the
fourth level.

One may construct a subgroup $SU(2)$ of the group $SU(n)$ 
for any 2D subspace of the effective Fock space. There are three such
"triads"  grouped in three vectors $\vec {\sf T}, \vec {\sf U},
\vec {\sf V}$ for the group $SU(3)$ with the symmetry operations acting
in the 3D space $\bar\Phi_3$ (cf. Ref. \onlinecite{ElDaub}). Adding fourth
dimension provides three more vectors $\vec {\sf W}, \vec {\sf Y},
\vec {\sf Z}$ representing the generators of the group $SU(4)$
together with the first three vectors.
\begin{eqnarray}\label{X.triad15}
&&{\sf T}^\pm = \frac{1}{2}(\lambda_1\pm i\lambda_2),~ {\sf T}_z = \lambda_3 \nonumber\\
&&{\sf U}^\pm = \frac{1}{2}(\lambda_6 \pm i\lambda_7),~
 {\sf U}_z = \frac{1}{2}(-\lambda_3 +\sqrt{3}\lambda_8)\nonumber\\
&&{\sf V}^\pm = \frac{1}{2}(\lambda_4 \pm i\lambda_5),~
 {\sf V}_z=\frac{1}{2}(\lambda_3
+\sqrt{3}\lambda_8), \nonumber\\
&& {\sf W}^{\pm} =
\frac{1}{2}\left(\lambda_9 \pm i\lambda_{10}\right),~
    {\sf W}_z = \frac{1}{2}\left(\lambda_3+\frac{1}{\sqrt{3}}\lambda_8+
    \frac{4}{\sqrt{6}}\lambda_{15}\right)\nonumber\\
&& {\sf Y}^{\pm} = \frac{1}{2}\left(\lambda_{11} \pm
i\lambda_{12}\right),~
    {\sf Y}_z = \frac{1}{2}\left(-\lambda_3+\frac{1}{\sqrt{3}}\lambda_8+
    \frac{4}{\sqrt{6}}\lambda_{15}\right)\nonumber\\
&& {\sf Z}^{\pm} = \frac{1}{2}\left(\lambda_{13} \pm
i\lambda_{14}\right),~
    {\sf Z}_z = \frac{1}{\sqrt{3}}\left(-\lambda_8+
    \sqrt{2}\lambda_{15}\right).
\end{eqnarray}

The original Hubbard operators $X^{\Lambda\Lambda'}$ are represented
via the Gell-Mann matrices in the basis $\bar\Phi_4$ (\ref{X.su10}) in
the following way:
\begin{eqnarray}\label{X.xsu4}
&& X^{\uparrow 0}=
\frac{1}{2}(\lambda_4+i\lambda_5),~~X^{0\uparrow} =\frac{1}{2}
(\lambda_4-i\lambda_5), \nonumber\\
&& X^{\downarrow 0}=
\frac{1}{2}(\lambda_6+i\lambda_7),~~X^{0\downarrow} =\frac{1}{2}
(\lambda_6-i\lambda_7)/2, \nonumber\\
&& X^{2\uparrow} = \frac{1}{2}(\lambda_9-i\lambda_{10}),~~
   X^{\uparrow 2}= \frac{1}{2}(\lambda_9+i\lambda_{10}) \nonumber\\
&& X^{2\downarrow} = \frac{1}{2}(\lambda_{11}-i\lambda_{12}),~~
   X^{\downarrow 2}= \frac{1}{2}(\lambda_{11}+i\lambda_{12})\nonumber \\
&& X^{\uparrow\downarrow} =\frac{1}{2}(\lambda_1 +
i\lambda_2),~~X^{\downarrow\uparrow}
=\frac{1}{2}(\lambda_1 -i\lambda_2) \nonumber\\
  &&X^{20}=\frac{1}{2}(\lambda_{13}-i\lambda_{14}),~~
  X^{02}=\frac{1}{2}(\lambda_{13}+i\lambda_{14})\nonumber\\
&& X^{\uparrow\uparrow} =\frac{1}{4}
\left(1+2\lambda_3+\frac{2}{\sqrt{3}}\lambda_8
+\frac{2}{\sqrt{6}}\lambda_{15}\right),\nonumber \\
&& X^{\downarrow\downarrow} =\frac{1}{4}
\left(1-2\lambda_3+\frac{2}{\sqrt{3}}\lambda_8
+\frac{2}{\sqrt{6}}\lambda_{15}\right),
\nonumber\\
  && X^{00}=\frac{1}{4}\left(1
-\frac{4}{\sqrt{3}}\lambda_8 +
\frac{2}{\sqrt{6}}\lambda_{15}\right), \nonumber\\
  &&  X^{22}= \frac{1}{4}(1-\sqrt{6}\lambda_{15})
\end{eqnarray}
The tree operators ${\sf T}^\pm, {\sf T}_z$ from the first triad
in the set (\ref{X.triad15})  describe the
spin-flip excitations in the homopolar subspace ${\cal N}=1$ of the
Hubbard atom. The three operators ${\sf Z}^\pm,
{\sf Z}_z$ from the last triad may be used in the description of excitations
in the two-particle sector ${\cal N}=\{0,2\}$ of the Hubbard and
Anderson model.
The operators forming the triads $\vec {\sf U}$ and $\vec {\sf V}$
intermix the states from the charge sectors ${\cal N}=0$ and
${\cal N}=1$, and the operators $\vec {\sf W}$ and $\vec {\sf Y}$
do the same for the sectors ${\cal N}=2,~{\cal N}=1$.

 In many physical applications the reduced
Anderson and Hubbard Hamiltonians with $U\to \infty$ are exploited. In this limit
the doubly occupied state $|2\rangle$ is completely suppressed. In the
appropriately reduced Fock space $\bar\Phi_3$ (\ref{X.su11})
possessing the $SU(3)$ symmetry the system (\ref{X.xsu4}) transforms
into
 \begin{eqnarray}\label{X.xsu43}
&& X^{\uparrow 0}=
\frac{1}{2}(\lambda_4+i\lambda_5),~~X^{0\uparrow} =\frac{1}{2}
(\lambda_4-i\lambda_5), \nonumber\\
&& X^{\downarrow 0}=
\frac{1}{2}(\lambda_6+i\lambda_7),~~X^{0\downarrow} =\frac{1}{2}
(\lambda_6-i\lambda_7)/2, \nonumber\\
&& X^{\uparrow\downarrow} =\frac{1}{2}(\lambda_1 +
i\lambda_2),~~X^{\downarrow\uparrow}
=\frac{1}{2}(\lambda_1 -i\lambda_2) \nonumber\\
 && X^{\uparrow\uparrow} =\frac{1}{2}
\left(\frac{2}{3}+\lambda_3+\frac{1}{\sqrt{3}}\lambda_8
\right),\nonumber \\
&& X^{\downarrow\downarrow} =\frac{1}{2}
\left(\frac{2}{3}-\lambda_3+\frac{1}{\sqrt{3}}\lambda_8 \right),
\nonumber\\
  &&  X^{00}=\frac{1}{3}\left(1
-\sqrt{3}\lambda_8 \right).
\end{eqnarray}

Within each triad the standard Pauli commutation relations
(\ref{3.comA}) for the components are valid. The commutation relations
between the operators from different subgroups are described by more
complicated structure factors.\cite{ElDaub} These relations in our
case may be derived from the general commutation relations (\ref{2.1b})
for the Hubbard operators (see the main text).

Two matrices used in the irreducible representation of the $SU(3)$ group are
\begin{equation}\label{apm}
{\sf T}_z =
 \left(
 \begin{array}{ccc}
 1 & 0 & 0  \\
 0 & -1 & 0  \\
 0 & 0 & 0 \\
 \end{array}
\right), ~~
  {\sf Q}=\frac{1}{3} \left(
 \begin{array}{ccc}
 1 & 0 & 0  \\
 0 & 1 & 0  \\
 0 & 0 & -2 \\
 \end{array}
\right)
\end{equation}
Their eigenvalues are
\begin{eqnarray}
M_T = 1,-1,~ 0;~~ M_Q = 1/3,~1/3, -2/3
\end{eqnarray}
for the states $|\uparrow\rangle, |\downarrow\rangle, |0\rangle$,
respectively.


\newpage
\begin{thebibliography}{99}
\bibitem{Nee62}Y. Ne'eman, Nucl. Phys. {\bf26}, 222 (1961)
\bibitem{GM64} M. Gell-Mann, Phys. Lett. {\bf8}, 214 (1964).
\bibitem{Zweig64} G. Zweig  CERN Reports No. 8181/Th 8419, 8419/Th 8412 (1964).
\bibitem{NeGm} M. Gell-Mann and Y. Ne'eman, {\it The Eightfold Way} (Westview Press 1964).
\bibitem{Barut64} A.O. Barut, Phys. Rev. {\bf135}, B839 (1964).
\bibitem{DGN65} Y. Dothan, M. Gell-Mann, and Y. Ne'eman, Phys. Lett. {\bf 17}, 148 (1965).
\bibitem{Engle} M.J. Englefield, {\it Group Theory and the Coulomb Problem}
    (Wiley, New York 1972).
\bibitem{MalMan79} I.A. Malkin and V.I. Man'ko, {\it Dynamical Symmetries and Coherent
    States of Quantum Systems} (Nauka, Moscow 1979), in Russian.
\bibitem{MORS65} N. Mukunda, L. O'Raifeartaigh, and E.C.G. Sudarshan,
    Phys. Rev. Lett. {\bf15}, 1041 (1965).
\bibitem{SMOR65} E.C.G. Sudarshan, N. Mukunda, and L. O'Raifeartaigh,
    Phys. Lett. {\bf 19}, 322 (1965).
\bibitem{MalMan65} I.A. Malkin and V.I. Man'ko, Sov.
    Phys. -- JETP Letters. {\bf 2}, 146 (1965); Sov. J. Nucl. Phys. {\bf 3}, 267 (1966).
\bibitem{GoLi59} S. Goshen and H.J. Lipkin, Annals Phys. {\bf 6}, 301, 310 (1959).
\bibitem{Barut65b} A.O. Barut, Phys. Rev. {\bf139}, B1433 (1965).
\bibitem{Hwa66} R.C. Hwa and J. Nuyts, Phys. Rev. {\bf 145}, 1188 (1966).
\bibitem{CowAus01} L.P. Kouwenhoven, D.G. Austing, and S. Tarucha, Rep. Progr. Phys., {\bf 64}, 701 (2001).
\bibitem{SaCi03} P.H. Sachrajda and M. Ciorga, {\it Nano-spintronics with lateral quantum dots}
    (Kluwer, Boston 2003).
\bibitem{CuFaR05} {\it Molecular electronics}, Lecture Notes in Physics, vol. 680
    (Eds. G. Cuniberti, G. Fagas, and K. Richter K, Springer, Berlin 2005).
\bibitem{Nat06} D. Natelson, {\it Handbook of Organic Electronics and Photonics}.
    (American Scientific Publishers, Valencia, CA 2006).
\bibitem{Hanson07} R. Hanson, L. P. Kouwenhoven, J. R. Petta, S. Tarucha, and
    L.M.K. Vandersypen, Rev. Mod. Phys. {\bf79}, 1217 (2007)
    \bibitem{Hub1} J. Hubbard, Proc. Roy. Soc. A {\bf 276}, 238 (1963).
\bibitem{Hub2} J. Hubbard, Proc. Roy. Soc. A {\bf 277}, 237 (1964).
\bibitem{Hub3} J. Hubbard, Proc. Roy. Soc. A {\bf 281}, 401 (1964).
\bibitem{Glara88}  L.I. Glazman and M.E. Raikh, Sov.
    Phys.-- JETP Lett. {\bf47}, 452 (1988).
\bibitem{NgLe88} T.K. Ng and P.A. Lee, Phys. Rev. Lett. {\bf 61}, 1768 (1988).
\bibitem{WTs81} A.M. Tsvelik and P.B. Wiegmann, Adv. Phys. {\bf32}, 453 (1983).
\bibitem{AFL81} N. Andrei, F. Furuya, and J.H. Loewenstein, Rev. Mod. Phys. {\bf55}, 331 (1983).
\bibitem{Anders67} P.W. Anderson, Phys. Rev. Lett. {\bf 18}, 1049 (1967).
\bibitem{And70} P.W. Anderson, J. Phys. C: Solid. State Phys. {\bf 3}, 2346 (1970).
\bibitem{AbMig70} A.A. Abrikosov and A.B. Migdal, J. Low Temp. Phys. {\bf 3}, 519 (1970).
\bibitem{FowZaw71} M. Fowler and A. Zawadowski, Solid State Commun. {\bf 9}, 471 (1971).
\bibitem{Jeff77} J.H. Jefferson, J. Phys. C: Solid. State Phys. {\bf 10}, 3589 (1977).
\bibitem{Hald78} F. D. M. Haldane, Phys. Rev. Lett. {\bf40}, 416 (1978).
\bibitem{GuiTaglia00} D. Giuliano and A. Tagliacozzo, Phys. Rev. Lett. {\bf 84}, 4677 (2000);
    D. Guiliano, B. Jouault, and A. Tagliacozzo, Phys. Rev. {\bf 63}, 125318 (2000).
\bibitem{EtoNazar00} M. Eto and Yu. Nazarov, Phys. Rev. Lett. {\bf85}, 1306 (2000); Phys. Rev. B {\bf64}, 085322 (2001).
\bibitem{KA01} K. Kikoin and Y. Avishai, Phys. Rev. Lett. {\bf86}, 2090 (2001);
    Phys. Rev. B {\bf65}, 115329 (2002).
\bibitem{KKA04} T. Kuzmenko, K. Kikoin, and Y. Avishai, Phys. Rev. Lett. {\bf89}, 156602 (2002);
 Phys. Rev. B {\bf 69}, 195109 (2004).
\bibitem{KAKr04} K. Kikoin, Y. Avishai and M.N. Kiselev. {\it  Dynamical symmetries in nanophysics}, in  "Nanophysics,
Nanoclusters and Nanodevices,"  (Nova Science Publishers, New York 2006) pp. 39-86.
\bibitem{SW66} J.R. Schrieffer and P.A. Wolff, Phys. Rev. {\bf 149}, 491 (1966).
\bibitem{ElDaub} J. Elliot and P. Dauber, {\it Symmetry in
Physics}, Ch.11 (Macmillan, London, 1979).
\bibitem{And75} P.W. Anderson, Phys. Rev. Lett. {\bf 34}, 953 (1975).
\bibitem{SchuF88} H.-B. Sch\"{u}ttler and A.J. Fedro, Phys. Rev. B {\bf 38}, 9063 (1988).
\bibitem{Taraph91} A. Taraphder and P. Coleman, Phys. Rev. Lett. {\bf 66}, 2814 (1991).
\bibitem{LaFi63} I.G. Lang and Yu. A. Firsov, Sov. Phys. JETP {\bf 16}, 1301 (1963).
\bibitem{Cornag95a} P.C. Cornaglia, D.R. Grempel, and H. Ness, Phys. Rev. B {\bf 71}, 075320 (2005).
\bibitem{Cornag95b} P.C. Cornaglia and D.R. Grempel, Phys. Rev. B {\bf 71}, 245326 (2005).
\bibitem{AlBrat03} A.S. Aleksandrov, A.M. Bratkovsky, and R.S. Williams, Phys. Rev. B {\bf 67}, 075301 (2003).
\bibitem{Koch06} J. Koch, M.E. Raikh, and F. von Oppen, Phys. Rev. Lett. {\bf96}, 056803 (2006).
\bibitem{Koch07} J. Koch, E. Sela, Y. Oreg, and F. von Oppen, Phys. Rev. B {\bf 75}, 195402 (2007).
\bibitem{LeWe09} M. Leijnse, M. R. Wegewijs, and M. H. Hettler, Phys. Rev. Lett. {\bf103}, 156803 (2009).
\bibitem{tilmaa} T. Tilma, M. Byrd, and E.C.G. Sudarshan. J. Phys. A: Math. Gen. {\bf 35}, 10445 (2002).
\bibitem{ReNe83} N. Read and D.M. Newns, J. Phys. C {\bf 16}, 3273 (1983).
\bibitem{Coleman84} P. Coleman, Phys. Rev. B {\bf 29}, 3035 (1984).
\bibitem{AfMars88} I. Affleck and B. Marston, Phys. Rev. B {\bf 37}, 3774 (1988).
\bibitem{KFO01} M.N. Kiselev, H. Feldmann, and R. Oppermann, Eur. Phys. J. B {\bf22}, 53 (2001).
\end{thebibliography}
\end{document}